Site-dependent hydrogenation on graphdiyne

P. A. S. Autreto[1], J. M. de Sousa[1] and D. S. Galvao[*1]

[1]Instituto de Física "Gleb Wataghin", Universidade Estadual de Campinas, Unicamp,
C.P. 6165, 13083-970, Campinas, São Paulo, Brazil.

Graphene is one of the most important materials in science today due to its unique and remarkable electronic, thermal and mechanical properties. However in its pristine state, graphene is a gapless semiconductor, what limits its use in transistor electronics. In part due to the revolution created by graphene in materials science, there is a renewed interest in other possible graphene-like two-dimensional structures. Examples of these structures are graphynes and graphdiynes, which are two-dimensional structures, composed of carbon atoms in $sp^2$ and sp-hybridized states. Graphdiynes (benzenoid rings connecting two acetylenic groups) were recently synthesized and some of them are intrinsically nonzero gap systems. These systems can be easily hydrogenated and the relative level of hydrogenation can be used to tune the band gap values. We have investigated, using fully reactive molecular dynamics (ReaxFF), the structural and dynamics aspects of the hydrogenation mechanisms of graphdiyne membranes. Our results showed that the hydrogen bindings have different atom incorporation rates and that the hydrogenation patterns change in time in a very complex way. The formation of correlated domains reported to hydrogenated graphene is no longer observed in graphdiyne cases.

---

[*] Corresponding author. Tel.: Fax: +55 19 35215373. Email address:
galvao@ifi.unicamp.br (D.S. Galvao).

1. INTRODUCTION

The three different hybridizations (sp, sp$^2$ and sp$^3$) of carbon (C) atoms allows the formation of many distinct allotropes, some of them only discovered in the last few decades: fullerenes [1], nanotube [2] and graphene [3], one of the most important subject in materials science today.

Graphene is a two-dimensional array of hexagonal sp$^2$ bonded carbon atoms. It has been theoretically investigated since late 1940's as a model to describe some properties of graphite [4,5]. Single layer graphite (graphene) was obtained from highly oriented pyrolytic graphite (HOPG) by Novoselov and Geim [3] using a "scotch tape" method. This discovery led to a revolution in materials science with an extraordinary number of theoretical and experimental works has published on the last decade. Although graphene presents several remarkable properties, there are some difficulties to be overcome before graphene-based nano electronics can become a reality. These difficulties are mainly related to its zero bandgap band structure, which precludes its direct use for some devices, such as digital transistors and diodes [6].

The advent of graphene created a renewed interest on some other 2D materials, as for example in the graphyne family [7,8]. Predicted by Baughman and co-workers in 1987 [8], graphyne is a generic name for a carbon allotrope family of 2D structures, where benzenoid rings are connected by acetylenic groups (Figure 1). One interesting aspect of these structures is the coexistence of sp and sp$^2$ hybridized carbon atoms. This family shares some of appealing graphene properties, such as the Dirac´s cone [3,7], but may exhibit intrinsically nonzero electronic gap systems [9,10]. As graphene can be considered the structural basis to create carbon nanotubes and fullerenes, graphyne can similarly generate nanotubes [11–13] and fullerenynes [14–16] (porous fullerenes).

In spite of years of efforts, only recently a member of this family was successfully synthesized: graphdiyne [17]. Graphdiyne can be described as a graphyne network, where diacetylenic groups (-C≡C-C≡C-) replace all acetylenic ones [8,16] (see Figure 1).

Previous studies have demonstrated that both graphene and graphdiyne exhibit extraordinary properties as high third-order nonlinear optical susceptibility, thermal resistance to conductivity, and through-sheet selective transport of ions [13,18]. Similarly to graphene [19,20], graphdiyne functionalization can be used for tuning the electronic gap. It can also directly modify the interaction of graphdiyne with the environment, which can be exploited in several technological applications [21,22]. Because of these properties, a considerable number of papers on pristine and functionalized graphynes and graphdiynes has been recently reported [16,23–27]. These results illustrate the remarkable variation in mechanical properties, fracture patterns, and exciting exotic electronic and geometrical properties exhibited by some hydrogenated graphyne and/or graphdiynes structures. Additionally, hydrogenation can be used to smartly tune their electronic band gaps [18]. With the exception of a recent work on the selective diffusion properties of hydrogen ($H_2$) on graphdiyne [28], most theoretical works addressing the hydrogenation of graphdiyne structures are based on ideally perfect models. These oversimplified models could not reflect the actual chemical processes involved in the complex hydrogenation dynamics. In order to obtain a more detailed description of these processes, more realistic (including aspects such as disorder) models are necessary. This is one of the objectives of our work. We have investigated, using fully reactive molecular dynamics, the structural and dynamical aspects of the atomic hydrogenation mechanisms of graphdiyne membranes at different conditions (temperatures, level of hydrogenation, etc.).

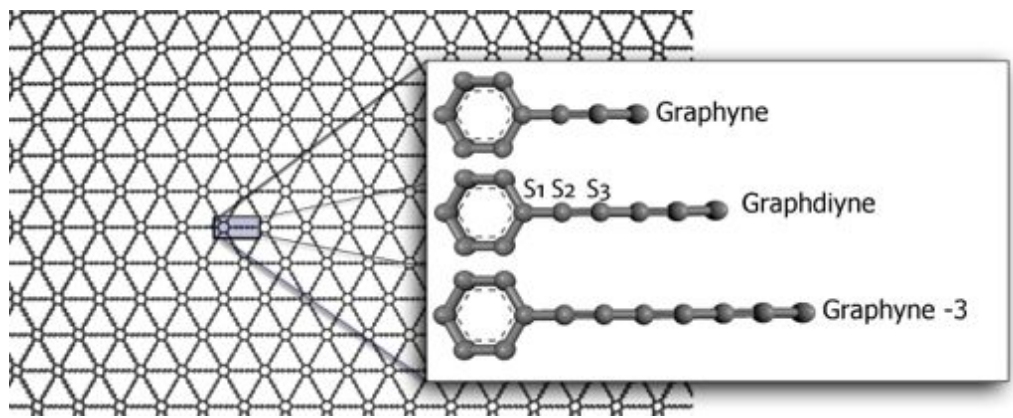

**Figure 1**. Structural model for α-graphyne structures.

## 2. METHODOLOGY

We have carried out molecular dynamics (MD) simulations in order to investigate the structural and dynamical aspects of the hydrogenation of graphdiyne membranes. The extensive MD simulations were performed using reactive force fields (ReaxFF) [29–31], as implemented in the Large-scale atomic/Molecular Massively Parallel Simulator (LAMMPS) code [32].

ReaxFF is a reactive force field developed by van Duin, Goddard III and co-workers for use in MD simulations. It allows simulations of many types of chemical reactions, including the formation of chemical bonds. It is similar to standard non-reactive force fields, like MM3 [29–31], where the system energy is divided into energy contributions associated with, amongst others, valence angle bending, bond stretching, and non-bonded van der Waals and Coulomb interactions. However, one main difference is that ReaxFF can handle bond formation and dissociation (making/breaking bonds) as a function of bond order values. ReaxFF was parameterized against DFT calculations, with the average deviations between the heats of formation predicted by the ReaxFF and the experiments equal to 2.8 and 2.9 kcal/mol, for non-conjugated and conjugated systems, respectively [29–31]. We have also calculated potential energy maps for a hydrogen atom placed at a distance of 1.5Å above the basal graphdiyne plane.

Our systems were composed of an isolated single-layer graphdiyne sheet (typical dimension of 185 Å x 150 Å (~6200 carbon atoms)) in contact with an atmosphere of atomic hydrogen atoms (~2500 H atoms). In order to speed up the simulations, we turn off the terms related to reactions between hydrogen atoms in ReaxFF force field, thus precluding H-H recombination during the MD runs. NVT ensemble, as implemented in LAMMPS code, was used and the typical time for a complete simulation run was 200 ps, with time-steps of 0.1 fs. Three different temperatures were considered: 300, 500 and 800 K. This methodology has been successfully applied to study hydrogenation processes in other similar 2D structures [19,20,33].

3. RESULTS AND DISCUSSION

Regarding chemical reactivity, graphdiyne has three distinct sites related to resonant (S1), single (S2) and triple bonds (S3) (Figure 1). These differences can be better visualized from 3D potential maps, which can also provide relevant information about which sites hydrogen atoms will preferentially bind to.

Potential maps for graphdiyne membranes and their hydrogenated forms are presented in Figure 2. There are significant differences among sites, in terms of preferential hydrogen binding. For the pristine graphdiynes (Figure 2(a)), the preferential binding sites are the carbon atoms at the triple bonds (S3). This is expected from a chemical point of view due to the higher local charge concentration [34]. A single hydrogen atom attached to one of these groups (Figure 2(b)) is enough to significantly modify the maps.

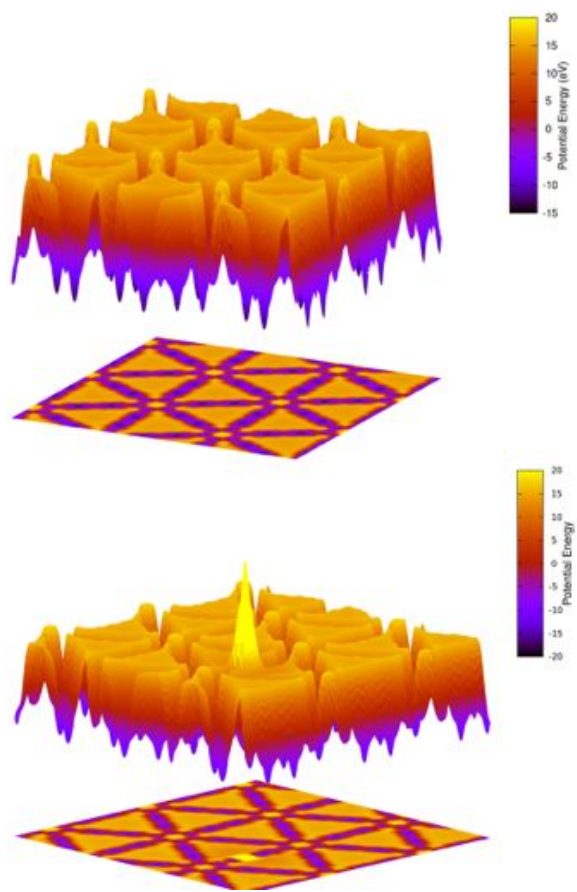

**Figure 2**. 3D potential energy maps (ReaxFF results) for the potential experienced by a hydrogen atom placed at a distance of 1.5Å above the basal graphdiyne plane. (a) graphdiyne and; (b) its hydrogenated form.

MD simulations corroborate the hydrogenation preference for site S3. Figure 3 presents the number of atoms bonded to different types of sites as a function of time, for three different temperatures (300, 500 and 800K). These results show that hydrogenation processes are highly dependent on temperature as well as on site type. For all temperatures considered here, the hydrogenation process occurs via fast absorption of hydrogen atoms, followed by a second stage presenting a linear regime of hydrogen bonding, until saturation. The order of most preferable sites to hydrogen biding is independent on temperature (S3, S2 and S1, respectively). The largest number of bonded sites occurs at 800K for S3, where 36% of S3 sites were hydrogenated. Although increasing temperatures facilitate the hydrogen binding, ordering and occupation fraction of hydrogenated sites are not affected. In Table 1 we present the fraction of hydrogenated sites, normalized to S1 values. We can from this table that there is only a small variation in the S2 and S3 values for the different temperatures.

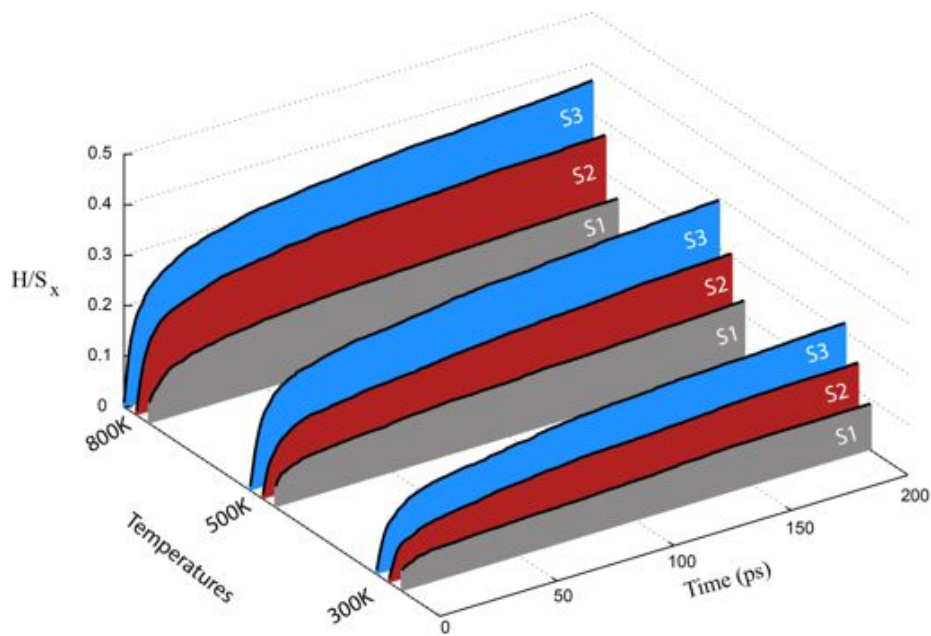

**Figure 3**. Incorporation of H atoms in time, for the sites indicated in Figure 1. Results for simulations at 300, 500 and 800K. See text for discussions.

Table 1. Fraction of bonded S2 and S3 sites normalized in relation to S1 values.

| Sites/Temperatures | 300K | 500K | 800K |
|---|---|---|---|
| **S2** | 1.6 | 1.5 | 1.6 |
| **S3** | 2.3 | 2.2 | 2.1 |

Another important result is that, in contrast to what was reported to the case of graphene hydrogenation [19], we did not observe the formation of correlated domains (islands of hydrogenated carbons). This it is a consequence of the porous graphdiyne structure, which allows larger out-of-plane deformations (in comparison to graphene). This results in an increase in the curvature and, consequently, in an increased local chemical reactivity. These results are also consistent with graphene fluorination data (fluorine is more reactive than hydrogen atoms) were the formation of these domains are also suppressed [20]. Even for the highest temperature considered here, 800 K, despite the significant number of bonded hydrogen atoms, the structural integrity of the membrane is not yet compromised, in contrast to what was observed in the graphene cases [19,20]. Our results also show that the hydrogenation saturation level is very far from 100% assumed in some idealized models [21,24,26,27]. This is consistent with the available experimental data [16] that show that high hydrogenation levels are very difficult to achieve. Our results illustrate the importance of considering large and disordered systems, and not only small ordered and fully hydrogenated models, which are unable to capture the complex hydrogenation dynamics.

Hydrogen binding to acetylenic groups can occur with different configurations and with different number of bonded sites. The number of these bonds per acetylenic linkers, as a function of time, is presented in Figure 4. In this figure we show the results for the cases for one hydrogen atom in diacetylene linker (O3, O2 and O1), as well as, for two hydrogen atoms (3-3', 2-3, 2-3', 1-2, 1-3) in the same linker. Three hydrogen atoms configurations is a rare event and occurred only for one case at 500K and 800K. We did not observe linkers with four, or even five hydrogen atoms in the same diacetylenic group.

At 300K, configurations with only one hydrogen atom in the same linker are the most common ones. A few number of 2H configurations are observed and are related to S3 and S2: 3-3', 2-3 and 2-3'. Similar results were obtained for the MD simulations at 500K, which favors the configuration of one atom bonded to site 3. In this temperature, 2 functionalized carbons atoms per chain became significant, and 3-3' H-configuration surpasses ones that include site 2. Actually, if you consider an isolated case, 3-3' should be the most common H-configuration, since site 3 is the most attractive site for hydrogen binding.

However, for 800K, when an expressive number of 1H-hydrogenation configurations occurs, an interesting result was observed. With few available free sites 3 in unbounded acetylenic groups, the acetylenic segments with just only bonded hydrogen atom at sites 2 became the most attractive place for the bonding of a second hydrogen atom at the 3 position. In this case, as we can infer from Figure 4, the number of 2-3 cases surpasses O1 and 3-3'ones. This is a direct consequence of the fact of first bonded atom in site 2 breaks one of the pi-bonds, thus favoring the hydrogen bonding of its neighboring atoms. It is interesting to notice that 3-3' is a common H configuration for low temperatures, while 2-3 configurations are the most common one for 800K. In this case, 3-3' H configurations become equivalent to 2-3' ones.

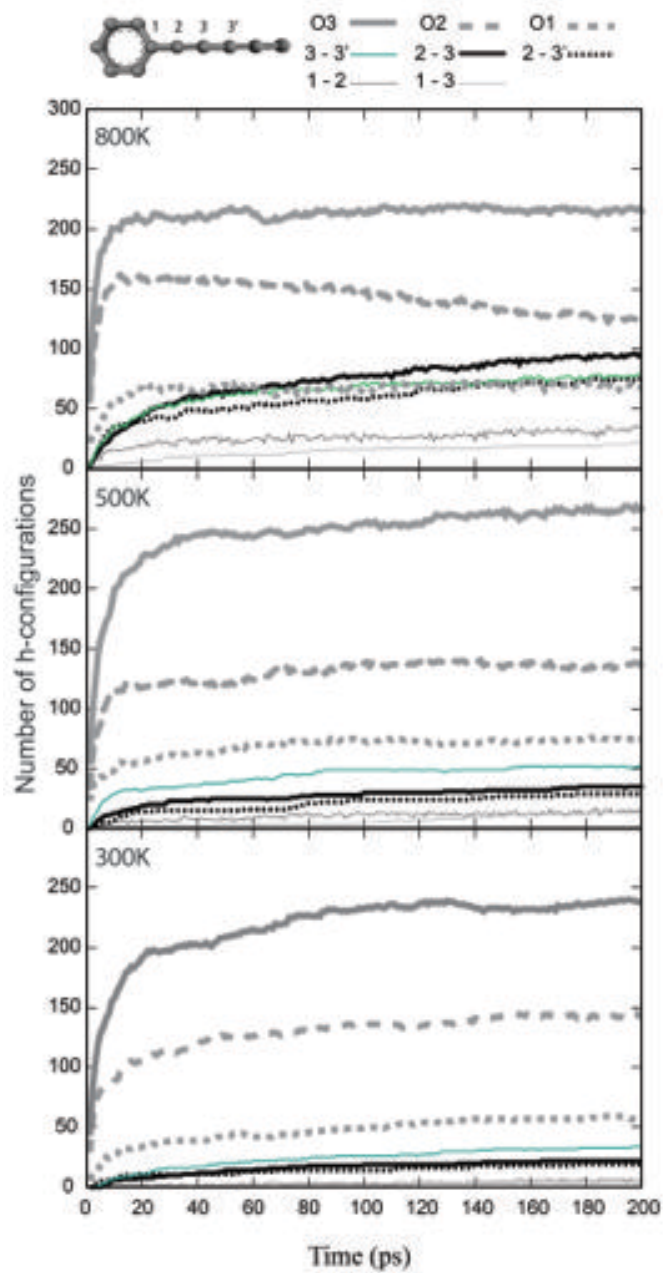

**Figure 4**. Number of configurations as a function of time for two hydrogen atoms bonded at the same diacetylene group for three different temperatures, 300K, 500K, and 800K. See text for discussions.

4. CONCLUSIONS

Here, we report a fully atomistic reactive molecular dynamics study of the dynamics of hydrogenation of graphdiyne membranes. We showed that the hydrogen bindings have different rate incorporations and that these rates change in time in very complex patterns. Initially, the most probable sites of hydrogenations are the carbon atoms forming the triple bonds, as chemically expected. But it changes in time and then the carbon atoms forming single bonds become the preferential biding sites. The formation of correlated hydrogenated domains observed in hydrogenated graphene is no longer observed into case of graphdiynes.

Acknowledgements

The authors would like to acknowledge Prof. A van Duin for helpful discussions. This work was supported in part by the Brazilian Agencies CAPES, CNPq and FAPESP. The authors thank the Center for Computational Engineering and Sciences at Unicamp for financial support through the FAPESP/CEPID Grant # 2013/08293-7.